# How humans evaluate AI systems for person detection in automatic train operation: Not all misses are alike

Romy Müller

*Faculty of Psychology, Chair of Engineering Psychology and Applied Cognitive Research, TUD Dresden University of Technology, Dresden, Germany*

Corresponding author:

Romy Müller

Chair of Engineering Psychology and Applied Cognitive Research

TUD Dresden University of Technology

Helmholtzstraße 10, 01069 Dresden, Germany

Email: romy.mueller@tu-dresden.de

Phone: +49 351 46335330

ORCID: 0000-0003-4750-7952

**Abstract**

If artificial intelligence (AI) is to be applied in safety-critical domains, its performance needs to be evaluated reliably. The present study aimed to understand how humans evaluate AI systems for person detection in automatic train operation. In three experiments, participants saw image sequences of people moving in the vicinity of railway tracks. A simulated AI had highlighted all detected people – sometimes correctly and sometimes not. Participants had to provide a numerical rating of the AI's performance and then verbally explain their rating. The experiments varied several factors that might influence human ratings: the types and plausibility of AI mistakes, the number of affected images, the number of people present in an image, the position of people relevant to the tracks, and the methods used to elicit human evaluations. While all these factors influenced human ratings, some effects were unexpected or deviated from normative standards. For instance, the factor with the strongest impact was people's position relative to the tracks, although participants had explicitly been instructed that the AI could not process such information. Taken together, the results suggest that humans may sometimes evaluate more than the AI's performance on the assigned task. Such mismatches between AI capabilities and human expectations should be taken into consideration when conducting safety audits of AI systems.

# 1. Introduction

How can we know whether an AI system is doing a good job? Reliable evaluations of AI performance are crucial in safety-critical domains such as medicine, industrial production, or autonomous driving. For instance, in automatic train operation (ATO), avoiding collisions or mitigating the severity of their consequences is a top priority. Although the assessment of collision risk requires far more than simply detecting potential obstacles (Müller & Schmidt, 2024), object detection is an important precondition for safe driving. Accordingly, a key focus of contemporary AI research in the automatic train operation context is to develop object detectors based on deep neural networks (DNN) (Cao et al., 2024; Ristić-Durrant et al., 2021). But how can we evaluate whether such systems are doing a good job? Different methods and standards for conducting safety audits of AI in the railway domain are being discussed (Gesmann-Nuissl & Kunitz, 2022; Roßbach et al., 2024). Still, it boils down to the fact that humans are responsible for evaluating the AI. Thus, it is essential to understand how they perform this task and what their evaluation outcomes depend on. These outcomes might vary with factors such as an AI mistake's type (e.g., miss, false alarm), its plausibility from a human perspective (e.g., missing an object that is hard to see vs. one that is in plain sight), its quantitative aspects (e.g., how often it occurs, how many objects are affected), and its potential consequences in terms of the danger emanating from a mistake (e.g., missing an object right on the tracks vs. at a safe distance). Moreover, methodological factors of conducting safety audits might play a role (e.g., how human evaluations are elicited).

The present study aimed to understand how humans evaluate the performance of AI systems for automatic train operation. Three experiments simulated the process of auditing an AI that detected people in dynamic image sequences and highlighted each detected person. The experiments varied the type and frequency of AI mistakes, the features of affected people and objects, and the methods used to elicit and integrate evaluation outcomes. The following sections will discuss why these factors might influence human evaluations of automated systems in general and AI in particular.

## 1.1 What factors shape human evaluations of AI performance?

### 1.1.1 AI accuracy and detection difficulty

An obvious influence on AI performance evaluations is the *accuracy of the AI's object detection*. If an AI misses an object, humans will presumably evaluate this AI worse than if it does not. Moreover, quantitative aspects of the problem are likely to play a role. Missing an object in only one or two images of a sequence might not be considered as bad as never detecting it, and missing one out of ten objects might not be considered as bad as missing the only object that is present in a scene.

That said, there is reason to assume that human evaluators do not simply count misses. Instead, their evaluations might depend on the *difficulty of detection and identification*. These tasks can be challenging for DNN in real-world scenarios, where not everything is as neat and tidy as in typical image datasets. Accordingly, the AI struggles with factors such as insufficient illumination, motion blur, small objects far away from the camera, occlusion, camouflage, and atypical perspectives (Beery et al., 2018). In the railway domain, detection systems have to deal with specific challenges such as small objects or bad weather (Cao et al., 2024). If such factors are present in images, this might make humans more forgiving of AI mistakes.

However, not all challenges faced by an AI are intuitive for humans. Although humans can often predict when and how the AI will misclassify images (Bos et al., 2019; Nartker et al., 2023), there also are puzzling AI mistakes that violate human expectations (Geirhos et al., 2020; Mueller, 2020). This plausibility of AI mistakes might shape human evaluations of AI performance. Humans seem to use their own image perception as a proxy to infer what an AI can see (Bos et al., 2019). Thus, they tend to distrust object detectors and underestimate their reliability when these systems fail on images that seem easy to humans (Madhavan et al., 2006). Accordingly, a previous study already reported that humans incorporate the plausibility of misclassifications into their evaluations of AI image classifiers (Heuer & Breiter, 2020). However, participants only had to provide holistic estimates of AI accuracy after seeing numerous classification instances, and then explain their strategies in a post-experimental interview. Thus, the findings cannot tell us how the plausibility of AI mistakes affects individual evaluation decisions.

*1.1.2 Type of AI mistake*

Aside from misses, AI systems can also produce false alarms, for instance when mistaking inanimate objects for people. False alarms were often found to have more detrimental impacts than misses on human performance and trust in automation (Dixon et al., 2007; Rice & McCarley, 2011; Wickens et al., 2005). Does this imply that the two types of mistakes differ in their inherent cognitive salience? Not necessarily, because most studies used difficult detection tasks, and the inferiority of false alarms might not generalise to situations in which humans are fully aware of all objects and AI mistakes. In fact, a study comparing easy misses and easy false alarms reported that they had similar impacts on human trust in the detection system (Madhavan et al., 2006). Another reason to doubt that false alarms will be evaluated worse than misses in the present study is their potential consequences. In automatic train operation, misses may cause collisions and collisions with people typically cause fatalities (Hampel et al., 2023). Conversely, false alarms could lead to unnecessary braking, which might only cause time delays. Thus, human evaluators might consider misses a greater problem.

Obviously, misses and false alarms are not mutually exclusive but can co-occur within one and the same situation. In this case, it is important to know how their effects on human evaluations combine. Multiple mistakes might dampen the impacts of each individual one, leading to under-additive effects. Alternatively, one might expect over-additive effects. For instance, in a fault detection task it was found that when the AI missed a target but at the same time falsely highlighted an unaffected area, the overall cost was higher than the sum of the costs for misses and false alarms alone (Müller et al., 2024). However, this study used a difficult detection task so that humans often failed to spot the miss. Moreover, it investigated human performance when assisted by an AI, instead of human evaluations of AI performance. These two measures can diverge, so that even when human performance is reduced by mistakes of an object detector, human trust in the system is not (Huegli et al., 2025). Thus, the generalisability of previous findings is questionable and it is important to study how misses, false alarms, and their combination will affect human evaluations of AI performance.

*1.1.3 Danger of the situation*

All previous considerations shared the implicit assumption that humans actually evaluate how the AI is performing its assigned task, and nothing but that. However, mismatches between AI capabilities and human expectations are common (Bach et al., 2024; Kocielnik et al., 2019). These expectations can be shaped by the knowledge, experiences, and attitudes humans hold about a particular domain. For instance, in the railway context it is common sense that collisions should be avoided whenever possible. Thus, humans might expect the AI to be sensitive to the danger of a situation (e.g., preferably detect people close to the tracks), and thus they might evaluate the AI more negatively if it misses people in danger. However, computer vision systems are fragmented, with each component specialising in a narrow subtask. One such subtask may be person detection. This is by far not enough to evaluate the risk of collision, for which train drivers use a wide range of cues such as a person's position, movement, visual appearance, inferred mental state, and a variety of context factors (Müller & Schmidt, 2024; Rosić et al., 2022). Still, AI components for person detection are needed and thus should be evaluated reliably. Accordingly, it is important to know whether such evaluations are subject to ill-calibrated human expectations.

*1.1.4 Method of eliciting human evaluations*

Finally, the method used to elicit human evaluations might play a role in shaping the outcomes. In principle, humans could evaluate AI performance for each individual image of a video sequence or for the sequence as a whole. Image-wise evaluations are more fine-grained and sequence-wise evaluations are more holistic. But do they also differ in their outcomes? For instance, sequence-wise evaluations could be more positive than the average evaluation of all images in the same sequence. This might occur, for instance, if evaluators do not consider it a major problem to miss a person for a few moments, given that this person was detected before and afterwards. Alternatively, sequence-wise evaluations could be more negative, assuming that a few bad images weigh more strongly than many good ones and thereby deteriorate the overall evaluation.

## 1.2 Present study

The aim of the present study was to find out how people evaluate the performance of AI systems for person detection in automatic train operation. Specifically, it was investigated how these evaluations are shaped by AI accuracy, detection difficulty, the type and quantity of AI mistakes, the danger of the situation, and the elicitation method. To this end, simulated AI audits were conducted in three experiments. Participants saw image sequences of dynamically unfolding scenes with people moving in the vicinity of the tracks. An AI had to detect and visually highlight all people. People were present in each image, but not every person was highlighted and not every highlight corresponded to a person. Participants had to evaluate whether the AI has done a good job. Note that throughout the study, no actual AI was used and all AI results were hand-crafted, ensuring a high level of experimental control over the influences to be tested. Experiment 1 aimed to establish what factors influence the evaluation of AI misses. Experiment 2 compared the results for misses and false alarms and investigated whether the presence of false alarms changes participants' evaluation of misses. Finally, Experiment 3 combined the two types of mistakes within the same sequences and tested whether their effects are additive.

All stimuli, instructions, rating data, syntax files, and explanation coding tables are made available via the Open Science Framework: https://osf.io/gyxbk/

## 2. Experiment 1

Experiment 1 asked what factors shape human evaluations of AI performance. First, it manipulated AI accuracy on three levels: perfect highlights, plausible misses, and implausible misses. The latter two differed in that it was easier versus harder to detect and identify the people missed by the AI. Second, the experiment tested whether evaluations depend on the number of images affected by a mistake and the number of people present in a sequence. Third, it assessed the influence of potential danger as operationalised by a missed person's position relative to the tracks. Finally, it varied whether human evaluations are elicited as a holistic rating of the entire sequence or numerically averaged over its individual images. It was hypothesised that ratings of AI performance are lower for implausible than plausible misses, lower when more images are affected, lower when fewer people are present, and lower when missed people are in danger. For the effects of elicitation method, no a priori hypotheses were formulated.

## 2.1 Methods

*2.1.1 Participants*

Thirty-three participants (21 female, 12 male) were recruited via the TUD Dresden University of Technology's participant pool. Their age ranged from 19 to 74 years ($M$ = 27.5, $SD$ = 13.3). All participants had normal or corrected-to-normal vision and were fluent in reading and writing German. They received partial course credit or a monetary compensation of 10 € per hour. All participants provided written informed consent and all procedures followed the principles of the Declaration of Helsinki.

*2.1.2 Apparatus and stimuli*

*Technical setup.* The experiments took place in a quiet lab room. Stimuli were presented on one of two desktop computers (screen size 24"), while a computer mouse and QWERTZ keyboard served as input devices. Written explanations of participants' ratings had to be entered on a separate laptop. The experiment was programmed with the Experiment Builder (SR Research, Ontario, Canada, Version 2.4.193).

*Instruction video.* Before starting the experiment, participants watched an instruction video of 2:36 minutes that was based on a Microsoft PowerPoint presentation. The video explained that the aim of the study was to investigate how people evaluate the safety of AI systems. It was made explicit that contemporary AI systems, including the present one, were *not* able to evaluate risk factors (e.g., whether people might enter the tracks). Instead, participants learned that the sole task of the AI was to detect whether people were present. The video explained that the AI had processed image sequences to detect people and that it had correctly decided that people were present somewhere in the sequence. However, the video also stated that it was unclear whether the AI had attended to all relevant areas, and that attended areas were presented in red. Based on a concrete image example, it was made clear that the AI should detect all people present and not mistake objects for people. Finally, the procedure of a trial was explained.

*Image sequences.* In total, 50 image sequences were shown, 2 of them during practice and 48 in the main experiment. Each sequence was composed of 10 consecutive static images. Each image depicted 1-15 people in the vicinity of railway tracks. Sequences were extracted from two rail-specific datasets, 9 from OSDaR23 (Tagiew et al., 2023) and 41 from RailGoerl24 (Tagiew et al., 2025). The OSDaR23 sequences depicted ordinary people performing ordinary actions at public railway stations (e.g., waiting or walking on the platform). The RailGoerl24 sequences depicted people in a shunting yard performing a wide variety of actions (e.g., crossing the tracks, lying on the tracks, interacting in groups).

The train was moving in 24.6 % of the sequences, while only the people were moving in the remaining 75.4 %. To capture this movement in a sequence of static images, individual images were manually picked from the video streams to ensure that there was a clear difference between consecutive images (e.g., people visibly changing their location or action). To achieve this, the time that had elapsed between two consecutive images varied between sequences.

To rule out confounds with stimulus-specific effects, similar scene contents had to be shown for each level of AI accuracy (i.e., perfect, plausible miss, implausible miss). To this end, the 48 sequences for the main experiment consisted of 16 triples. Each triple had a common theme (e.g., person lying on the tracks while the train is approaching, people with umbrellas in a group of several people) and consisted of three exemplars (e.g., three different people lying on the tracks). These exemplars were distributed across three parallel sets of 16 sequences, one from each triple. Finally, each set was assigned to a particular AI accuracy level for a particular participant, and this assignment was counterbalanced across participants.

*Highlights.* To indicate which people the AI had detected, highlights were presented in the form of transparent red ellipses overlayed on each detected person. These highlights covered all body parts but intentionally lacked clear boundaries or colour gradients to prevent participants from evaluating low-level visual features (e.g., spatial precision, distribution of colour across a person's body).

Three types of sequences varied AI accuracy as reflected by the highlights. For *perfect* AI, each person in each image was highlighted. For *plausible misses*, one or more people were not highlighted that were hard to see. In half of the sequences, this difficulty was a matter of perception (i.e., small size or low in contrast) and in the other half it was a matter of identification (i.e., partial occlusion or unusual posture). For *implausible misses*, the missed people were clearly visible (i.e., in plain sight, sufficiently large, normal posture). It is important to note that plausible and implausible misses were relative rather than absolute categories. That is, a miss was considered plausible when the affected person was harder to see than in other images of the same sequence or than other people within the same image. Accordingly, the results only provide information about *more or less* visibility within the same situation, not about absolute criteria or psychophysical boundaries.

**Figure 1.** Stimulus examples. Original images (top row) and the same images with highlights (bottom row). Only one image per sequence is shown, while in the experiment participants saw 10 consecutive images extracted from a video.

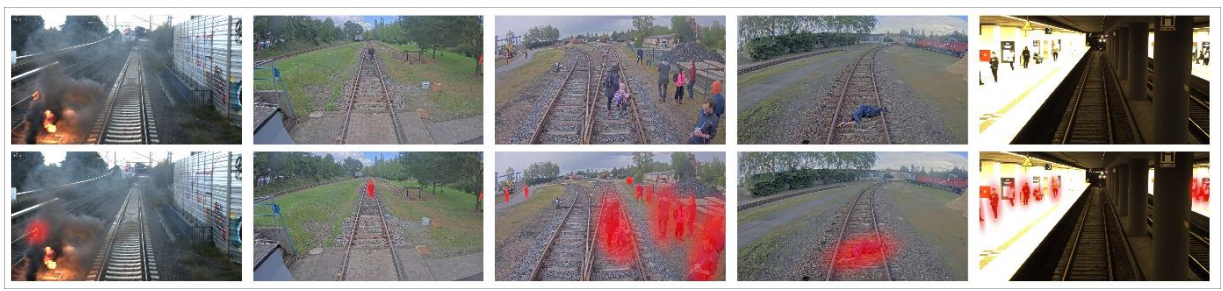

*Screens.* The experimental screens contained images, interaction elements, and text in white font on a black background. All text was presented in German. Six types of screens were shown in each trial: transition screens, sequence preview screens, image rating screens, sequence replay screens, sequence rating screens, and explanation screens (see Figure 2). *Transition screens* enabled participants to enter the next phase of a trial by pressing a button (e.g., start sequence preview). Ten consecutive *sequence preview screens* presented the original images without highlights in full screen mode (1920 × 1080 px). Ten consecutive *image rating screens* presented two images (960 × 540 px) next to each other in the upper part of the screen. The original image on the left-hand side was labelled "Original image" and the image with highlights on the right-hand side was labelled "This is what the AI has attended to." Below the images, a text asked "How well has the AI done its job?" Below this question, there was a slider with the poles "very poorly" and "very well". Participants could click at any position, providing an integer value between 0 and 100. Ten consecutive *sequence replay screens* again presented the two images (i.e., original and highlighted), while the rest of the screen was empty. The *sequence rating screen* showed ten light green squares (representing the sequence) as well as a slider. Finally, *explanation screens* allowed participants to review the images while providing a written explanation of their rating. In addition to the images there were two buttons (labelled "previous" and "next") which could be used to skip back and forth between the images of a sequence. Moreover, a numerical sequence identifier and the current image number were shown. At the bottom of the screen there was a button saying "Finished entering the explanation."

*Table for explanations*. A Microsoft Excel sheet was provided for participants to explain their ratings. One column with the heading "Sequence" contained the identifiers for the 48 sequences in the randomised order in which they were shown to the respective participant. The second column with the heading "Explanation" was empty and had to be filled in by participants.

*2.1.3 Procedure*

An overview of the procedure is provided in Figure 2. After watching the instruction video, participants completed two practice trials and 48 experimental trials, each presenting one sequence. The order of sequences was randomised for each participant individually. After starting a trial by pressing a button on a transition screen, participants saw a sequence preview consisting of ten consecutive images, each being shown for 500 ms. After the last image of the preview, another transition screen appeared on which participants could press a button to start the image rating procedure. For each individual image, they had to answer the question "How well has the AI done its job?" by selecting a value on the rating scale. Clicking the respective slider position automatically brought them to the next image. If they wanted to go back to a previous image, they could do so by pressing the "B" key on their keyboard. This allowed them to redo their ratings for as many images as they wanted. After rating all ten images of a sequence, participants were brought to another transition screen and could press a button to start their rating of the sequence as a whole. This brough them to an automatic replay of the entire sequence, with each pair of original and highlighted image being shown for 1000 ms. After the tenth image, the sequence rating screen appeared, and participants could enter a rating by clicking a position on the scale. This brought them to a transition screen on which they could press one of two buttons, one to enter the explanation screens and click through the sequence again, and one to directly indicate that they had finished their written explanation (i.e., if they felt no need to revisit the images). After 48 trials, the experiment ended. Overall, it took between 45 minutes and two-and-a-half hours.

**Figure 2.** Procedure of a trial. Participants saw a sequence preview, provided individual ratings of the AI for each individual image, saw a replay of the sequence, provided a holistic rating of the AI for the entire sequence, and explained their rating.

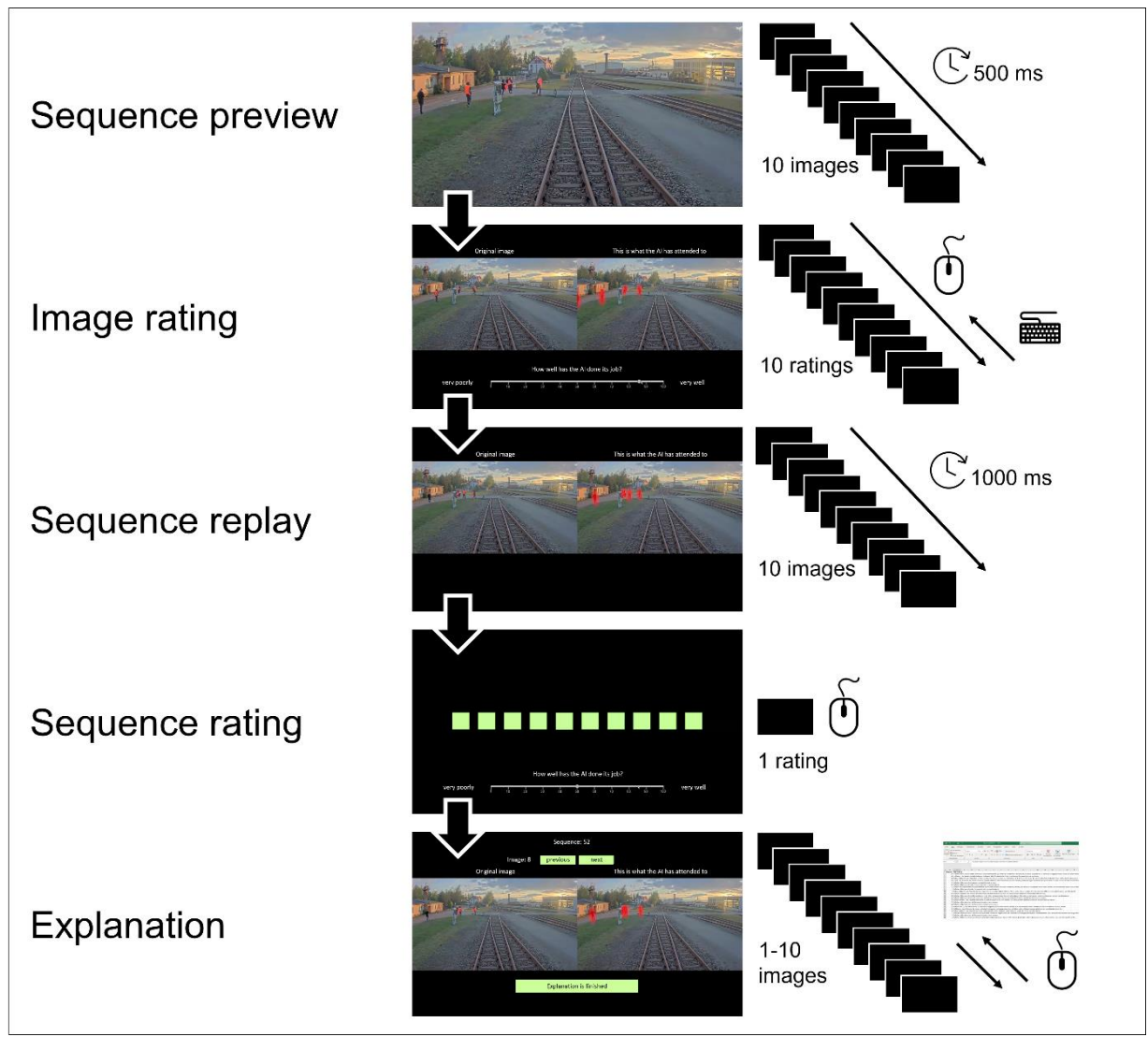

*2.1.4 Data analysis*

*Ratings of AI performance*. The analyses described below were conducted on participants' holistic ratings of the sequence as a whole. Individual image-wise ratings were only used to assess the effects of elicitation method (see below). To statistically analyse what factors influence evaluations, analyses of variance (ANOVAs) were computed. The first was a one-way repeated-measures ANOVA with the 3-level factor *AI accuracy (perfect, plausible miss, implausible miss)*. All subsequent analyses determined how AI accuracy interacted with the remaining factors (i.e., detection difficulty, number of affected images, number of people, people's position relative to the danger zone, elicitation method). Thus, one potential influence per analysis was included as a second repeated-measures factor and analysed for its interaction with AI accuracy. The specific analyses and deviations from this procedure will be described in the respective parts of the Results section. If sphericity was violated, a Greenhouse-Geisser correction was applied and the degrees of freedom were adjusted accordingly. To determine statistical significance, an alpha level of $p < .05$ was used and all pairwise comparison were performed with Bonferroni correction.

*Explanations of ratings*. In addition to participants' AI performance ratings, their written explanations of these ratings were analysed. To this end, the information elements contained in each explanation were assigned to a number of categories. These categories were determined iteratively, based on a qualitative content analysis of the data. For statements about people, the categorisation started with the categories that train drivers use to evaluate the risk of collision (Müller & Schmidt, 2024). Over the course of the analysis, these were refined and adjusted to fit the present study's written explanations. Overall, 44 categories were used (see Table 1).

**Table 1.** Topics, subtopics, and categories used to code participants' written explanations of their AI performance ratings.

| Topics | Subtopics | Categories |
|---|---|---|
| AI performance | Detection | Correct detection, miss, correct rejection, false alarm |
| | Number of people | All people, number of people |
| | Frequency and time | Always, never, number of images, image ID, point in time, loss and fluctuation |
| | AI capabilities | Interdependence, comparison, inference |
| Image context | Consequences | Danger, relevance |
| | Difficulty | Own perception, perceivability, confusability, light and weather, occlusion, parts visible |
| People | Type | Identity |
| | Location | On tracks, distance to tracks, distance to train, position in image, relation to objects, relation to people |
| | Behaviour | Movement, action, approach or leave tracks, approach or leave train, potential action |
| | Features | Size and shape, posture, clothing and colour, gadgets |
| Objects | Detection | Object should be detected as person, attached to person, dangerous, correctly rejected, falsely detected |

The overarching topics of these categories were AI performance, image-related factors, people, and objects. First, categories summarised under *AI performance* were coded when an explanation referred to the success or failure of person detection (e.g., the AI has detected, has missed, has not falsely identified), when the explanation quantified the people that were detected or missed (e.g., all people, most people, three people), when it mentioned the frequency and time of detection (e.g., the AI has never detected the person, missed the person in three images, detected the person too late, lost the person in image 5), and when the explanation referred to the AI's actual or desirable capabilities (e.g., the AI was able to detect smaller people in other images, the AI should know that occluded people are still there). Second, categories subsumed under *image context* included all statements specific to the current image sequence, except for those concerning the people and objects to be detected. These statements referred to the consequences of an AI mistake (e.g., the AI has missed the person in most danger, detecting this person would be relevant) and to the difficulty of detection (e.g., the AI has seen the person earlier than I did, the person was hard to see, the person disappeared behind a tree, only a foot was visible). Third, statements about *people* included information about people's identity (e.g., child, worker), location (e.g., on the tracks, close to the tracks, far in the background, next to the car), behaviour (e.g., moving, playing, approaching the tracks) and physical features such as posture and clothing (e.g., the person is small, kneeling, wearing a safety vest). Finally, references to *objects* included all information about objects the AI should be able to detect, has falsely detected, or has correctly rejected (e.g., the pram should be detected, umbrellas should be considered a part of the person, the AI has falsely detected a traffic sign, the AI has not mistaken the safety vest for a person).

For each explanation, all categories were coded once if the explanation referred to the respective content. Subsequently, for each participant and category the percentage of trials was computed in which the category was used. These percentages were analysed qualitatively to infer what contents play a prominent role in participants' ratings of AI performance.

## 2.2 Results and discussion

### 2.2.1 Ratings of AI performance

The following sections will report how participants' ratings of AI performance depended on AI accuracy and how this interacted with potential influencing factors. An overview of all mean values and standard deviations is provided in Table 2.

**Table 2.** Means and standard deviations (in parentheses) for participants' ratings of AI performance in Experiment 1.

|  |  | Perfect | Plausible miss | Implausible miss |
|---|---|---|---|---|
| AI accuracy |  | 95.0 (4.4) | 66.2 (13.6) | 53.9 (15.5) |
| Detection difficulty | Low | - | 56.4 (17.5) | - |
|  | Intermediate | - | 73.2 (15.0) | - |
|  | High | - | 69.9 (16.0) | - |
| Number of affected images | 2 | - | 73.4 (15.0) | 64.3 (17.1) |
|  | 3 | - | 65.6 (16.8) | 45.3 (20.0) |
|  | 4-9 | - | 63.6 (14.7) | 48.6 (16.4) |
|  | 10 | - | 61.3 (16.8) | 54.3 (18.2) |
| Number of people | 1 | 95.0 (8.2) | 57.4 (19.5) | 46.4 (17.8) |
|  | 2-4 | 94.3 (6.9) | 72.6 (13.5) | 50.8 (17.9) |
|  | 5-7 | 97.0 (3.3) | 75.9 (15.5) | 64.1 (17.6) |
|  | 8+ | 95.7 (5.8) | 58.7 (18.2) | 55.3 (20.8) |
| Position relative to danger zone | Inside, close | - | 36.3 (27.9) | 44.9 (18.2) |
|  | Inside, medium | - | 56.9 (20.9) | 45.7 (19.0) |
|  | Inside, far | - | 62.3 (15.6) |  |
|  | Outside, close | - | 68.6 (13.3) | 63.7 (18.4) |
|  | Outside, medium | - | 76.5 (16.4) | 66.9 (15.0) |
|  | Outside, far | - | 81.9 (13.1) |  |
| Elicitation method | Sequence-wise | 95.0 (4.4) | 66.2 (13.6) | 53.9 (15.5) |
|  | Image-wise | 95.4 (4.5) | 74.5 (8.9) | 69.0 (8.9) |

*AI accuracy.* The one-way ANOVA testing for effects of *AI accuracy* (*perfect, plausible miss, implausible miss*) yielded a significant effect, $F(1.7, 53.1) = 212.612$, $p < .001$, $\eta_p^2 = .869$. Participants' ratings were highest in case of perfect AI (95.0), considerably lower for plausible misses (66.2), and again lower for implausible misses (53.0), all $ps < .001$ (see Figure 3A). Numerically, the drop between perfect AI and plausible misses (28.8) was more than twice as large as that between plausible and implausible misses (12.3). Thus, the AI was punished even for missing people that were hard to perceive or identify, and the additional cost of missing something obvious was much smaller.

**Figure 3.** Mean ratings of AI performance depending on AI accuracy. (A) Experiment 1, (B) Experiment 2, (C) Experiment 3. Error bars represent standard errors of the mean, stars indicate which comparisons were statistically significant.

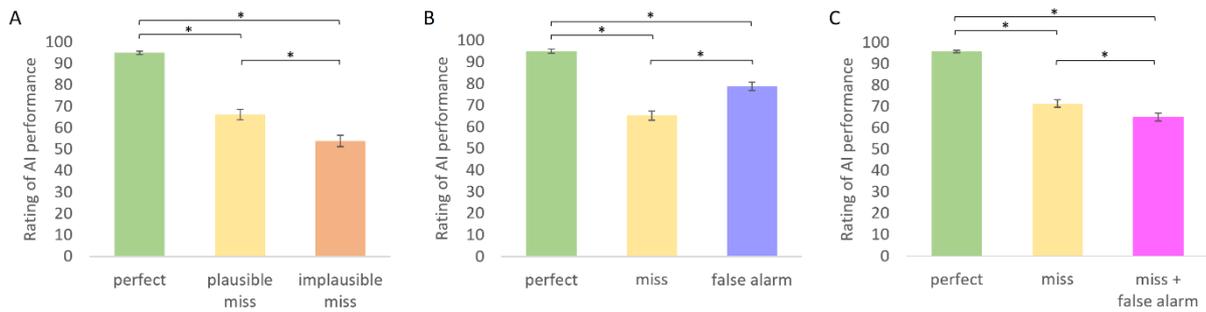

A possible reason for this pronounced punishment of plausible misses might be that they were not so plausible after all. For humans, it is rather easy to identify people even when they are in unusual postures or partially occluded. Thus, they might have expected the AI to have no problems with that, either. Therefore, a control analysis was performed to check whether ratings of plausible misses depended on the difficulty of detecting the missed people. To this end, a one-way ANOVA with the repeated-measures factor *detection difficulty* (*low, intermediate, high*) was conducted for plausible misses only. A main effect of detection difficulty, $F(1.5, 48.2) = 22.345$, $p < .001$, $\eta_p^2 = .411$, indicated that ratings were indeed lower when detection was easy (56.4) than when it was intermediate or difficult (73.2 and 69.9, respectively), both $p$s $< .001$, while the latter two did not differ from each other, $p = .305$. These numerical values suggest that although detection difficulty played a role, it cannot explain the substantial cost for plausible misses: even when the missed people were barely visible, participants still only gave ratings of around 70, compared to 95 when the AI's detection performance was perfect.

*Number of affected images.* To test how participants' ratings depended on quantitative aspects of the AI mistake, a 2 (*AI accuracy: plausible miss, implausible miss*) x 4 (*number of affected images: 2, 3, 4-9, 10*) repeated-measures ANOVA was conducted. These four levels were chosen in order to achieve a balanced number of trials in each cell. This is because there were many more sequences in which people were either missed only briefly (2 or 3 images) or missed all the time (10 images), but only few sequences in which they were missed in a medium number of images. The ANOVA revealed a main effect of AI accuracy, $F(1, 32) = 62.888$, $p < .001$, $\eta_p^2 = .663$, a main effect of number of affected images, $F(3, 96) = 24.921$, $p < .001$, $\eta_p^2 = .438$, as well as an interaction of both factors, $F(2.3, 73.7) = 6.664$, $p = .001$, $\eta_p^2 = .172$ (see Figure 4A).

**Figure 4.** Mean ratings of AI performance depending on AI accuracy and the number of affected images. (A) Experiment 1, (B) Experiment 2, (C) Experiment 3. Error bars represent standard errors of the mean, stars indicate which comparisons were statistically significant.

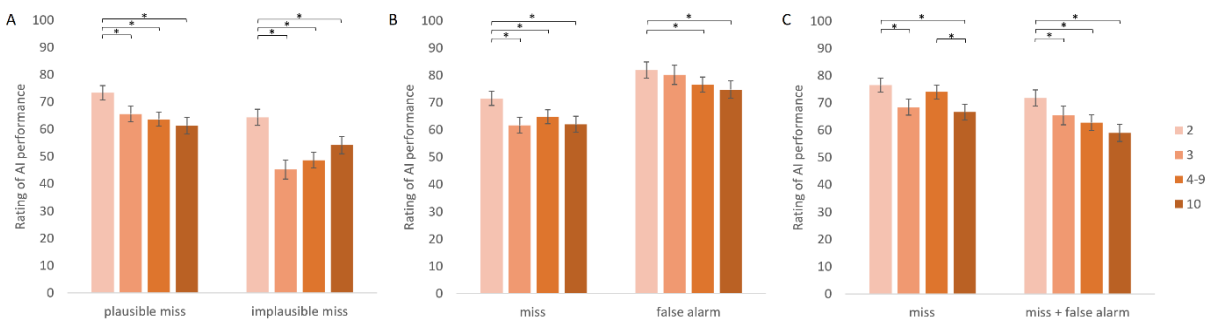

Overall, the AI was rated worse when misses were more frequent. Interestingly, only the difference between two images (68.9) and all higher numbers was significant, all *p*s < .001. Conversely, when more than two images were affected, it did not matter whether it was 3, 4-9, or 10 (55.5, 56.1, and 57.8, respectively), all *p*s > .9. Despite the significant interaction, this pattern of results was found both for plausible and implausible misses. In both cases, the only significant difference was that between two and more than two images. Thus, missing people was generally considered less problematic when it only occurred very briefly.

*Number of people*. To investigate how ratings depended on the people present in a sequence, a 3 (*AI accuracy: perfect, plausible miss, implausible miss*) x 4 (*number of people: 1, 2-4, 5-7, 8+*) repeated-measures ANOVA was conducted. It revealed a main effect of AI accuracy, $F(2,64) = 182.298$, $p < .001$, $\eta_p^2 = .851$, a main effect of number of people, $F(3,96) = 23.286$, $p < .001$, $\eta_p^2 = .421$, and an interaction of both factors, $F(6,192) = 13.144$, $p = .001$, $\eta_p^2 = .291$ (see Figure 5A). Ratings were lowest when only one person was present, and then increased with up to seven people, all *p*s < .001. Conversely, the difference between 2-4 and 5-7 people was not significant, $p > .7$. However, there was unexpected drop in ratings for large groups of eight or more people, $p < .001$, and this pattern was observed both for plausible and implausible misses. This surprising finding will be picked up in the General Discussion.

**Figure 5.** Mean ratings of AI performance depending on AI accuracy and the number of people present in a sequence. (A) Experiment 1, (B) Experiment 2, (C) Experiment 3. Error bars represent standard errors of the mean, stars indicate which comparisons were statistically significant.

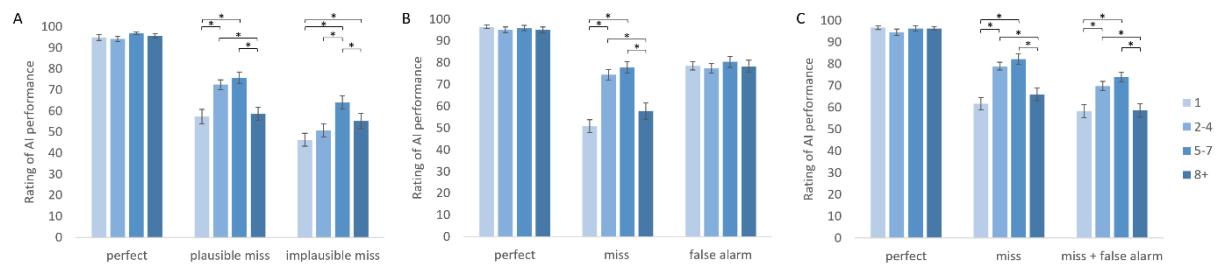

*Position relative to the danger zone*. To investigate how participants' ratings depend on the danger resulting from a miss, the position of missed people was determined. Danger depends on two aspects of position: whether missed people are inside the train's profile (i.e., area in which they would be hit if the train was passing) and how far they are away from the train (i.e., camera). People were considered close if the distance was less than 40 meters, medium between 40 and 80 meters, and far if the distance exceeded 80 meters. Due to a systematic scarcity of far-away people for implausible misses (see below), the analysis had to be carried out separately for the two types of miss.

For plausible misses, a one-way repeated-measures ANOVA was conducted with the 6-level factor *position relative to danger zone* (*inside close, inside medium, inside far, outside close, outside medium, outside far*). There was a significant effect of position relative to the danger zone, $F(2.8, 89.6) = 45.180$, $p < .001$, $\eta_p^2 = .585$ (see Figure 6A, also for the significance of pairwise comparisons). Ratings were lower when the missed people were inside the train's profile or closer. By far the worst ratings were given for sequences in which the AI missed people on the tracks right in front of the train, for instance because they were occluded or in an unusual posture (36.3). In contrast, missing people far away and outside the train's profile led to rather positive ratings (81.9). The large difference between these two values (45.6) suggests that people's position was the influencing factor with the strongest impact on participants' ratings overall.

For implausible misses, the missed people tended to be closer and rarely were located far in the distance. Therefore, the data were collapsed for medium and far distances, and a one-way repeated-measures ANOVA was conducted with the 4-level factor *position relative to danger zone* (*inside close, inside further, outside close, outside further*). There was a main effect of position relative to the danger zone, $F(2.1, 66.3) = 41.020$, $p < .001$, $\eta_p^2 = .562$. However, the pairwise comparisons indicated that ratings were lower when the missed people were inside the profile than when they were outside, while distance did not play a significant role (see Figure 6A).

**Figure 6.** Mean ratings of AI performance depending on AI accuracy and the position of missed people or falsely detected objects relative to the danger zone. (A) Experiment 1, (B) Experiment 2, (C) Experiment 3. Error bars represent standard errors of the mean, stars indicate which comparisons were statistically significant.

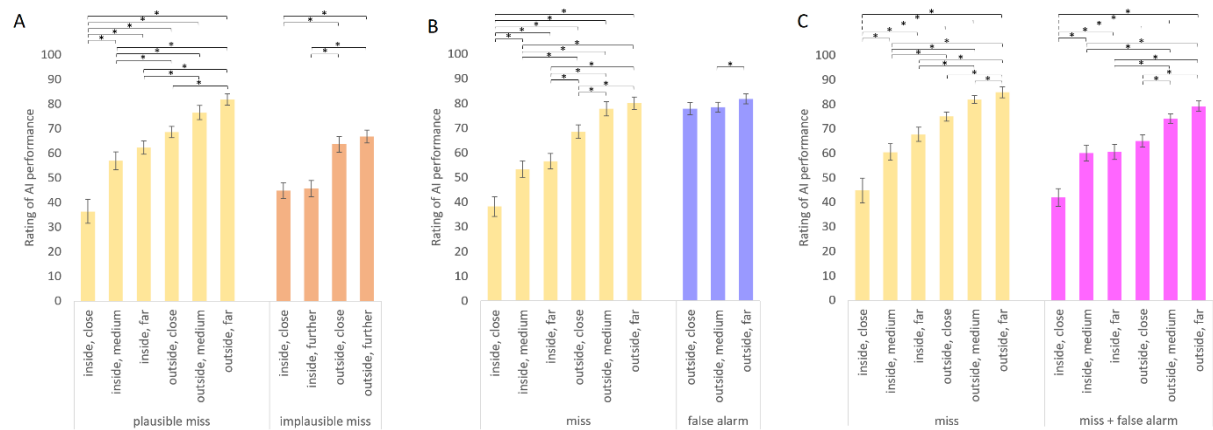

*Elicitation method*. To test for influences of elicitation method, participants' holistic ratings for the entire sequence were compared to the average of their ratings for the 10 individual images of the same sequence. A 3 (*AI accuracy: perfect, plausible miss, implausible miss*) x 2 (*elicitation method: sequence-wise, image-wise*) repeated-measures ANOVA was conducted. It revealed a main effect of AI accuracy, $F(2,64) = 261.340$, $p < .001$, $\eta_p^2 = .891$, a main effect of elicitation method, $F(1,32) = 72.279$, $p < .001$, $\eta_p^2 = .693$, and an interaction of both factors, $F(2,64) = 61.358$, $p = .001$, $\eta_p^2 = .657$ (see Figure 7A). This interaction indicated that only in case of perfect AI, it did not matter how ratings were elicited. Conversely, for both plausible and implausible misses, participants rated the overall sequence worse than its individual images, both $p$s < .001. This sequence cost was numerically higher for implausible misses (15.1) than for plausible misses (8.3).

**Figure 7.** Mean ratings of AI performance depending on AI accuracy and the method used to elicit participants' ratings. (A) Experiment 1, (B) Experiment 2, (C) Experiment 3. Error bars represent standard errors of the mean, stars indicate which comparisons were statistically significant.

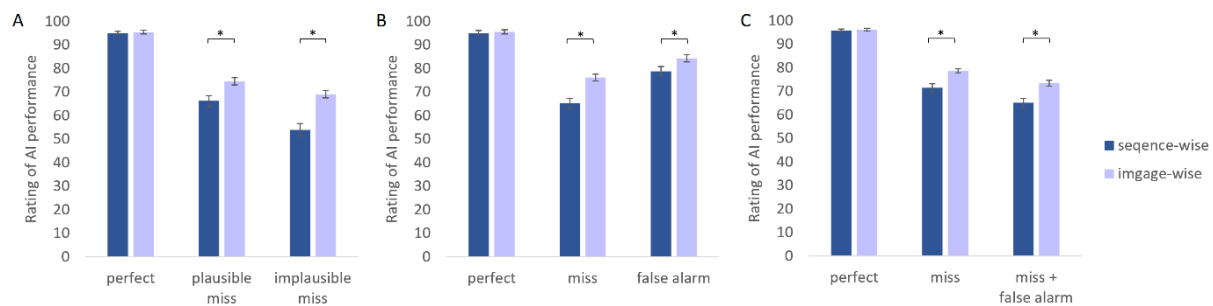

*2.2.2 Explanation of ratings*

In addition to the quantitative analysis of AI performance ratings, participants' explanations of these ratings were analysed qualitatively. To this end, the concepts mentioned in their explanations were assigned to 44 categories. The percentages of concepts in each category are presented in Figure 8A and Table 3. The following section will highlight the most relevant aspects that become apparent when inspecting the figure.

**Figure 8.** Percentage of trials in which participants used a particular concept in the written explanation of their rating. Error bars represent standard errors of the mean.

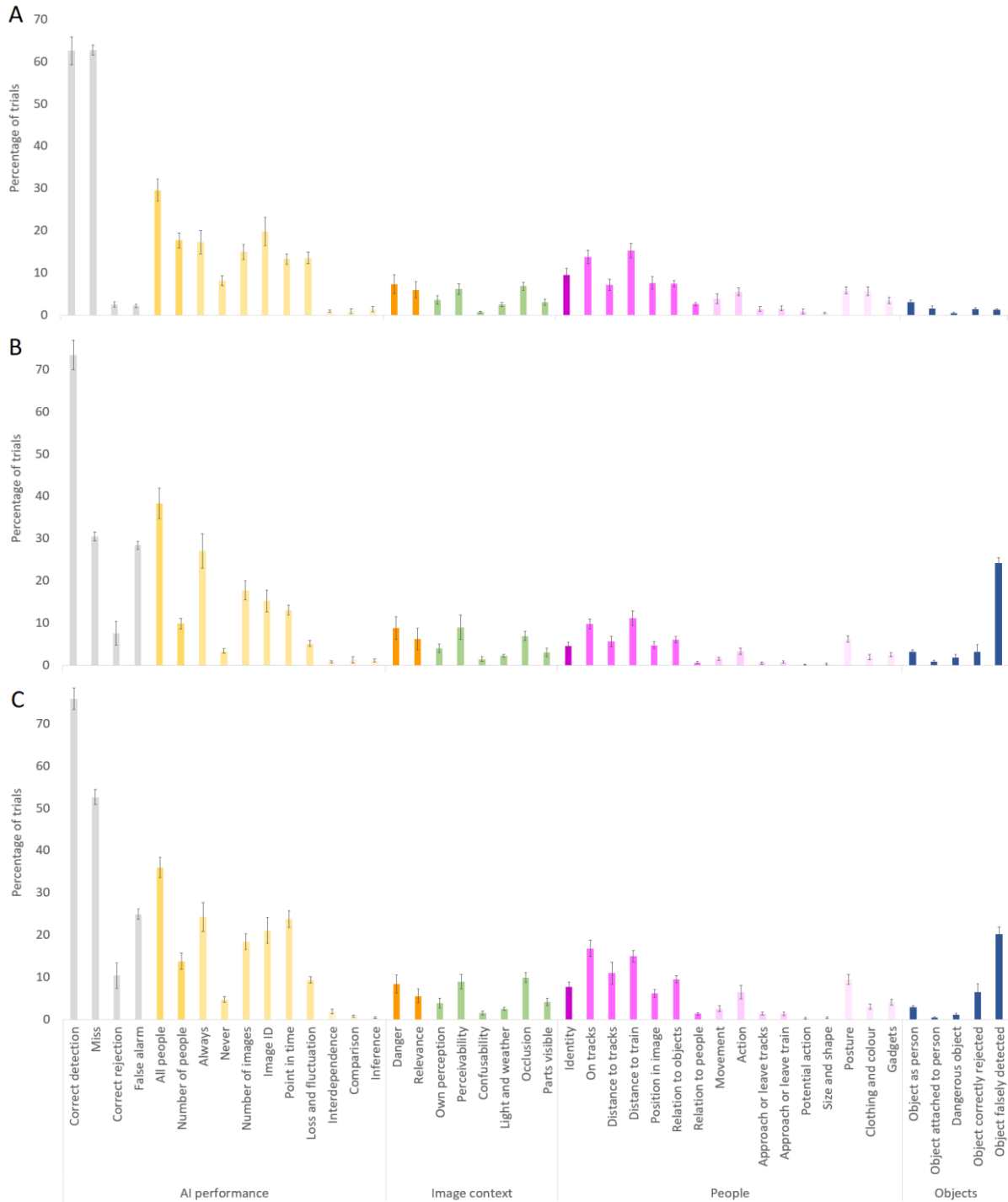

**Table 3.** Mean percentage of concepts and standard deviation (in parentheses) in each explanation category for Experiments 1-3.

| Topic | Category | Experiment 1 | Experiment 2 | Experiment 3 |
|---|---|---|---|---|
| AI performance | Correct detection | 62.5 (19.1) | 73.4 (20.1) | 75.8 (14.5) |
| | Miss | 62.7 (7.0) | 30.4 (6.0) | 52.6 (10.0) |
| | Correct rejection | 2.5 (3.5) | 7.6 (15.9) | 10.4 (17.4) |
| | False alarm | 2.1 (2.3) | 28.3 (5.3) | 24.9 (7.0) |
| | All people | 29.5 (15.1) | 38.3 (21.0) | 36.0 (13.9) |
| | Number of people | 17.7 (10.4) | 9.8 (7.3) | 13.8 (11.1) |
| | Always | 17.2 (15.9) | 27.0 (23.5) | 24.2 (19.9) |
| | Never | 8.1 (6.5) | 3.3 (3.0) | 4.7 (3.8) |
| | Number of images | 14.9 (10.5) | 17.7 (13.0) | 18.4 (10.8) |
| | Image ID | 19.8 (19.3) | 15.2 (15.0) | 21.0 (17.4) |
| | Point in time | 13.3 (7.2) | 13.0 (6.7) | 23.7 (11.7) |
| | Loss and fluctuation | 13.5 (7.9) | 5.1 (3.9) | 9.3 (4.2) |
| | Interdependence | 0.9 (1.8) | 0.8 (1.6) | 1.8 (3.0) |
| | Comparison | 0.9 (3.1) | 1.3 (4.1) | 0.8 (1.6) |
| | Inference | 1.4 (3.5) | 1.1 (2.0) | 0.4 (1.4) |
| Image context | Danger | 7.3 (13.1) | 8.8 (15.5) | 8.4 (12.3) |
| | Relevance | 5.9 (11.0) | 6.2 (14.4) | 5.6 (9.3) |
| | Own perception | 3.6 (5.8) | 4.0 (5.7) | 3.9 (6.4) |
| | Perceivability | 6.1 (7.2) | 9.0 (16.4) | 9.0 (9.9) |
| | Confusability | 0.7 (1.4) | 1.5 (2.9) | 1.5 (3.0) |
| | Light and weather | 2.5 (2.6) | 2.2 (2.2) | 2.5 (2.0) |
| | Occlusion | 6.8 (5.7) | 6.9 (6.2) | 9.9 (6.5) |
| | Parts visible | 3.0 (4.1) | 3.1 (6.0) | 4.1 (5.0) |
| People | Identity | 9.5 (9.4) | 4.5 (4.8) | 7.7 (6.6) |
| | On tracks | 13.7 (8.9) | 9.7 (6.7) | 16.8 (11.0) |
| | Distance to tracks | 7.1 (7.5) | 5.6 (7.0) | 10.9 (14.8) |
| | Distance to train | 15.2 (10.0) | 11.2 (10.0) | 14.9 (8.1) |
| | Position in image | 7.6 (8.6) | 4.7 (4.6) | 6.1 (5.8) |
| | Relation to objects | 7.4 (4.5) | 6.0 (4.4) | 9.5 (5.4) |
| | Relation to people | 2.6 (2.5) | 0.6 (1.2) | 1.3 (1.9) |
| | Movement | 3.9 (6.6) | 1.5 (2.2) | 2.5 (4.1) |
| | Action | 5.5 (4.9) | 3.3 (4.3) | 6.5 (9.0) |
| | Approach or leave tracks | 1.5 (3.2) | 0.5 (1.3) | 1.4 (2.1) |
| | Approach or leave train | 1.6 (3.1) | 0.8 (1.8) | 1.3 (2.0) |
| | Potential action | 0.9 (3.6) | 0.1 (0.5) | 0.3 (0.9) |
| | Size and shape | 0.5 (1.0) | 0.3 (1.1) | 0.4 (1.1) |
| | Posture | 5.8 (4.8) | 6.2 (4.4) | 9.5 (6.7) |
| | Clothing and colour | 5.6 (6.0) | 1.9 (3.6) | 3.0 (3.4) |
| | Gadgets | 3.5 (4.3) | 2.5 (2.6) | 4.1 (4.0) |
| Objects | Object as person | 3.1 (2.4) | 3.2 (2.6) | 2.8 (2.7) |
| | Object attached to person | 1.5 (3.8) | 0.8 (2.1) | 0.4 (0.9) |
| | Dangerous object | 0.4 (1.1) | 1.8 (4.1) | 1.1 (2.4) |
| | Object correctly rejected | 1.3 (2.3) | 3.2 (9.4) | 6.4 (11.4) |
| | Object falsely detected | 1.2 (1.7) | 24.1 (7.5) | 20.1 (10.0) |

*AI performance*. Most explanations contained information about AI accuracy. The first two grey bars in Figure 8A (correct detection, miss) are the highest of all, suggesting that participants usually base their ratings on the AI detecting or not detecting people. This is not surprising but can be interpreted as a manipulation check, showing that participants actually performed their task. What seems more interesting are the third and fourth grey bars (correct rejection, false alarm). In Experiment 1, no false alarms occurred. The fact that this rarely played a role in participants' explanations might suggest that they did not consider the AI's specificity as a positive influence on their evaluations.

The yellow bars describe how often and how reliably the AI detected or missed people. The categories that seem most interesting here are those describing the immediacy (point in time, e.g., "the person was detected too late") and consistency of detection (loss and fluctuation, e.g., "the AI lost the person when it moved further away"). These categories were featured in about 13 % of participants' explanations and thus seemed important. Conversely, the low height of the next three bars (interdependence, comparison, inference) suggests that participants mostly evaluated the AI's actual task instead of reasoning about capabilities that the AI presumably has, does not have, or should have.

*Image context*. Many explanations referred to image features that made detection either important or difficult. The two orange bars (danger, relevance) suggest that ratings depended not only on the success of detection but also on its consequences. The next three bars (own perception, perceivability, confusability) refer to detection difficulty. While participants often referred to their own perception and to people's perceivability, they rarely mentioned confusability with objects. This is in line with the finding that ratings were higher when detection was difficult, while unusual postures did not seem to count as an excuse for missing a person. The next three bars (light and weather, occlusion, parts visible) specify the contextual factors underlying detection difficulty. Among them, occlusion was mentioned most often. Overall, these results suggest that participants partly based their ratings on how hard it was for the AI to perform its task.

*People*. Among the concepts referring to people, the most frequently used ones were about position, especially whether people were on the tracks and how far they were away from the train. In fact, these were the most frequently used image-related (as opposed to AI-related) categories of all. Interestingly, concepts about people's behaviour (movement, action, approach) were much rarer, and participants rarely mentioned what people might be doing next (potential action). Thus, the dynamics of sequences seemed to play a smaller role than position. Similarly, people's physical features (posture, clothing, gadgets) were mentioned much less than position. Overall, these explanation results corroborate the finding that participants' numerical ratings were most strongly affected by a missed person's position relative to the danger zone.

*Objects*. Explanations rarely contained information about objects. If they did, this mostly occurred when participants concluded that an object (e.g., car, pram) should be detected as a person because it might contain a person.

*2.2.3 Interim summary*

Experiment 1 revealed several influences on participants' evaluations of AI performance. Ratings were lower when people were missed that are easy to detect and identify, when misses affected more images of a sequence, when missed people were the only ones present in a scene, when they were in danger, and when ratings were elicited holistically for the entire sequence. These findings were corroborated by participants' explanations, revealing that they paid attention to task difficulty, the number and timing of misses and, most importantly, to people's position.

Yet, Experiment 1 left an important question unanswered: how participants' evaluations of misses depend on the context of other AI mistakes. Specifically, it is unclear whether misses would be rated differently if the AI was prone to false alarms. The relevance of this question is highlighted by a peculiar observation, namely that the risk of false alarms did not seem to play a role: Participants did not punish the AI for detecting people that were barely visible (e.g., tiny black dot in the distance). Quite the contrary, they praised the AI for its superior detection capabilities, even highlighting people that they could not possibly see, yet. A possible reason is that no false alarms occurred throughout the experiment, reducing the need for specificity. This possibility was investigated in Experiment 2.

# 3. Experiment 2

The aims of Experiment 2 were threefold. First, it attempted to replicate the influences on evaluations observed in Experiment 1, thereby assessing their stability. Second, it investigated the effects of false alarms as a different type of AI mistake. Given the pronounced impacts of danger in Experiment 1, it was expected that false alarms will receive less negative ratings than misses as they presumably pose less danger. Third, Experiment 2 asked how the results for plausible misses depend on the context of AI performance. Will people evaluate misses differently when false alarms can occur? There are three opposing hypotheses. First, participants might rate plausible misses more positively than in Experiment 1, namely when the increased saliency of false alarms leads them to value specificity. Following the same line of reasoning, the opposite outcome might be expected for perfect AI. That is, some sequences might receive lower ratings when the AI (correctly) highlights people that are impossible to identify. Instead of following the assumption that more is always better, participants might reward more realistic AI performance. The second hypothesis assumes that contrast effects might lead to the opposite result: Participants might rate plausible misses lower when the alternative mistake is not more problematic (like implausible misses) but less problematic (like false alarms). Finally, if mistake context is irrelevant, plausible misses and perfect AI should receive ratings similar to those obtained in Experiment 1.

## 3.1 Methods

*3.1.1 Participants*

Thirty-three new participants (24 female, 9 male) were recruited via the TUD Dresden University of Technology's participant pool. Their age ranged from 18 to 67 years ($M$ = 28.0, $SD$ = 9.8). All participants had normal or corrected-to-normal vision and were fluent in reading and writing German. They received partial course credit or a monetary compensation of 10 € per hour. Participants provided written informed consent and all procedures followed the principles of the Declaration of Helsinki.

*3.1.2 Apparatus and stimuli*

The stimuli were identical to those used in Experiment 1, except for the placement of highlights: Instead of implausible misses, false alarms were used in one third of the sequences. That is, all people were highlighted correctly, but in addition to that, one inanimate object was highlighted (e.g., safety vests, posters with people, parts of the railway infrastructure). To facilitate comparisons between misses and false alarms, the latter appeared on the very same images that were used for plausible misses in the respective AI accuracy condition. In this way, the two types of AI mistake did not differ in terms of frequency or image characteristics. An exception was made for two sequences, where a falsely highlighted warning vest was not visible in the images that had been used for plausible misses.

*3.1.3 Procedure*

The procedure was identical to that of Experiment 1, with the only difference that the factor *AI accuracy* had the three levels *perfect*, *miss* (identical to the plausible misses of Experiment 1), and *false alarm*.

*3.1.4 Data analysis*

The methods used to analyse the data were mostly identical to those of Experiment 1. Exceptions will be made explicit in the respective parts of the Results section.

## 3.2 Results and discussion

*3.2.1 Ratings of AI performance*

An overview of all mean values and standard deviations is provided in Table 4.

**Table 4.** Means and standard deviations (in parentheses) for participants' ratings of AI performance in Experiment 2.

|  |  | Perfect | Miss | False alarm |
|---|---|---|---|---|
| AI accuracy |  | 95.1 (5.8) | 65.3 (12.0) | 78.8 (11.3) |
| Number of affected images | 2 | - | 71.5 (13.0) | 81.9 (10.6) |
|  | 3 | - | 61.7 (16.9) | 80.1 (12.0) |
|  | 4-9 | - | 64.8 (12.7) | 76.6 (14.4) |
|  | 10 | - | 62.1 (16.7) | 74.8 (13.8) |
| Number of people | 1 | 96.5 (5.4) | 50.9 (16.8) | 78.6 (11.3) |
|  | 2-4 | 95.1 (7.4) | 74.5 (14.2) | 77.5 (12.4) |
|  | 5-7 | 95.9 (6.8) | 77.9 (14.5) | 80.5 (14.6) |
|  | 8+ | 95.1 (7.5) | 57.8 (21.4) | 78.4 (16.2) |
| Position relative to danger zone | Inside, close | - | 38.3 (23.1) | - |
|  | Inside, medium | - | 53.4 (19.2) | - |
|  | Inside, far | - | 56.6 (17.7) | - |
|  | Outside, close | - | 68.5 (15.8) | 77.8 (14.2) |
|  | Outside, medium | - | 77.8 (16.1) | 78.6 (11.3) |
|  | Outside, far | - | 80.1 (14.7) | 81.8 (11.8) |
| Elicitation method | Sequence-wise | 95.1 (5.8) | 65.3 (12.0) | 78.8 (11.3) |
|  | Image-wise | 95.5 (5.4) | 76.1 (8.4) | 84.3 (8.6) |

*AI accuracy*. There was a highly significant effect of AI accuracy, $F(2,64) = 136.378$, $p < .001$, $\eta_p^2 = .810$. Just like in Experiment 1, ratings were highest in case of perfect AI (95.1) and much lower for (plausible) misses (65.3). False alarms were in between, but were clearly considered less problematic than misses (78.8), all $p$s $< .001$ (see Figure 3B).

*Number of affected images*. The ANOVA revealed main effects of AI accuracy, $F(1,32) = 44.142$, $p < .001$, $\eta_p^2 = .580$, and number of affected images, $F(3,96) = 11.433$, $p < .001$, $\eta_p^2 = .263$, while the interaction missed the significance level, $F(3,96) = 2.586$, $p = .058$, $\eta_p^2 = .075$ (see Figure 4B). Overall, the AI was rated worse when mistakes occurred in more images. However, just like in Experiment 1, only the difference between two images (76.7) and all higher numbers of images was significant, all *p*s < .001. Conversely, when more than two images were affected, it did not matter whether it was 3, 4-9, or 10 (70.9, 70.7, and 68.4, respectively), all *p*s > .8. This pattern was similar for misses and false alarms.

*Number of people*. The ANOVA revealed main effects of AI accuracy, $F(2,64) = 130.354$, $p < .001$, $\eta_p^2 = .803$, and number of people, $F(3,96) = 21.501$, $p < .001$, $\eta_p^2 = .402$, as well as an interaction of both factors, $F(4.0, 127.1) = 21.682$, $p = .001$, $\eta_p^2 = .404$ (see Figure 5B). For misses, the results mirrored those of Experiment 1: Ratings were low when the only existing person was missed, increased with more people being present, both *p*s < .001, but it did not matter how many people were present, $p > .9$. However, for large groups of eight or more people, ratings dropped, $p < .001$. For perfect AI and false alarms, the number of people present in a sequence did not have any effect, all *p*s > .7.

*Position relative to the danger zone*. For misses, there was a significant effect of position relative to the danger zone, $F(2.9, 91.7) = 38.968$, $p < .001$, $\eta_p^2 = .549$ (see Figure 6B). Ratings were lower when the missed people were inside the train's profile or closer to the train. By far the lowest ratings were given when the AI missed people on the tracks right in front of the train (38.3). In contrast, missing people far away and outside the train's profile led to rather positive ratings (80.1). Just like in Experiment 1, the difference between these two values (41.8) was the numerically largest influence observed in the entire experiment. For false alarms, only two of the 48 sequences contained an object in the train's profile. Thus, the analysis was only performed for objects outside the profile but varying in their distance to the train, using a one-way repeated-measures ANOVA with the 3-level factor *position relative to danger zone* (*outside close, outside medium, outside far*). This analysis yielded a much smaller but yet significant main effect of position relative to the danger zone, $F(2,64) = 3.963$, $p = .024$, $\eta_p^2 = .110$. Ratings were slightly higher for large than medium distances, $p = .041$.

These data reveal another interesting aspect. Overall, misses were rated lower than false alarms. However, an inspection of Figure 6B suggests that the low ratings for misses were primarily given for objects inside the danger zone, which did not exist for false alarms. To test whether the inferiority of misses persists when controlling for confounds of position, a 2 (*AI accuracy: miss, false alarm*) x 3 (*position relative to danger zone: outside close, outside medium, outside far*) repeated-measures ANOVA was computed. There still was a significant but now much smaller main effect of AI accuracy, $F(1,32) = 5.540$, $p = .025$, $\eta_p^2 = .148$, as well as a main effect of position relative to the danger zone, $F(2,64) = 14.719$, $p < .001$, $\eta_p^2 = .315$, and an interaction, $F(2,64) = 4.967$, $p = .010$, $\eta_p^2 = .134$. This interaction was due to the fact that misses were rated lower than false alarms only when objects were close, $p < .001$, but not when the distance was medium or large, both *p*s > .4.

*Elicitation method*. When comparing sequence- and image-wise elicitation, there were main effects of AI accuracy, $F(2,64) = 135.362$, $p < .001$, $\eta_p^2 = .809$, and elicitation method, $F(1,32) = 132.957$, $p < .001$, $\eta_p^2 = .806$, as well as an interaction of both factors, $F(2,64) = 54.180$, $p = .001$, $\eta_p^2 = .629$ (see Figure 7B). This interaction indicated that only for perfect AI, it did not matter how ratings were elicited. Conversely, for misses and false alarms, participants rated the overall sequence worse than its individual images, both *p*s < .001. This sequence cost was twice as large for misses (10.8) as it was for false alarms (5.4).

*3.2.2 Explanations of ratings*

The percentages of concepts in each category used to code participants' explanations are presented in Figure 8B and Table 3. The pattern of results closely mirrored Experiment 1, particularly for categories describing images and people. Again, participants frequently considered the consequences of detection (danger, relevance), detection difficulty (own perception, perceivability, occlusion) and the position of people (on tracks, distance to train). However, there also were a few noticeable differences to Experiment 1, which will be described in the following paragraphs.

*AI performance*. Misses were mentioned only half as often as in Experiment 1, while false alarms were now mentioned as often as misses. This is not surprising given that half of the misses were replaced by false alarms in the sequences. In line with this, the numbers of detected or missed people were also mentioned less often. Moreover, participants referred to correct rejections somewhat more often, suggesting that the presence of false alarms shifted their focus and made it more salient when the AI did not make this type of mistake.

*Image context*. Interestingly, the image context was not considered any less than in Experiment 1. For the consequences of detection, this seemed to be because false alarms also triggered participants to comment on danger and relevance, only in the opposite direction (i.e., claiming that false alarms were not as dangerous and relevant as misses). For detection difficulty, the similarity to Experiment 1 might stem from the fact that difficult sequences (i.e., plausible misses) were retained in the stimulus set.

*People and Objects*. Participants mentioned people less often overall, but the pattern of category frequencies was closely mirrored that of Experiment 1. The overall reduction seems reasonable given that for false alarm sequences, describing people was less relevant. The close match of the category distribution speaks to the stability of participants' reasoning processes across experiments. Not surprisingly, statements mentioning inanimate objects were greatly enhanced as half of the AI mistakes now highlighted these objects.

*3.2.3 Interim summary*

Experiment 2 revealed three main findings. First, it closely replicated the effects of the influencing factors on ratings and explanations, attesting a high stability of the findings. Second, false alarms led to more positive evaluations of AI performance overall, but not when the affected objects were far away. Thus, missed people were only considered more problematic when they were in acute danger of being hit by the train. Third, the fault context (i.e., false alarms rather than implausible misses) did not noticeably affect the results for plausible misses or perfect detections. Thus, the risk of mistaking objects for people did not seem to make participants more conservative and placing higher value on an AI's specificity. This independence from the context of other faults is surprising and thus was investigated further in Experiment 3, where misses and false alarms occurred in the same images.

# 4. Experiment 3

Experiment 3 combined misses and false alarms within the same sequences and compared this combination to pure misses. This was done for two reasons. First, it served to establish whether the independence of miss ratings from fault context in Experiment 2 was a more general phenomenon. If so, the results for misses observed in Experiments 1 and 2 should once again be replicated. If not, one might expect misses to receive higher ratings in Experiment 3, where the alternative was a much more severe AI problem. Second, Experiment 3 investigated how the effects of misses and false alarms combine. If they additively affect human ratings, the cost in terms of lower ratings for combined mistakes (miss plus false alarm) should be lower than that for pure misses by a value similar to the cost for pure false alarms in Experiment 2. Conversely, if the effects get dampened, the deviation should be lower – that is, the whole should be less than the sum of its parts.

## 4.1 Methods

*4.1.1 Participants*

Thirty-three new participants (23 female, 10 male) were recruited via the TUD Dresden University of Technology's participant pool. Their age ranged from 18 to 68 years ($M$ = 28.8, $SD$ = 13.0). All participants had normal or corrected-to-normal vision and were fluent in reading and writing German. They received partial course credit or a monetary compensation of 10 € per hour. Participants provided written informed consent and all procedures followed the principles of the Declaration of Helsinki.

*4.1.2 Apparatus and stimuli*

The experimental setup and stimuli were identical to those in the previous experiments with only one exception: the AI accuracy condition *misses + false alarms* combined the plausible misses from Experiments 1 and 2 with the false alarms from Experiment 2 in the same sequences.

*4.1.3 Procedure*

The procedure was identical to that of Experiments 1 and 2. Only the experimental design differed in that the within-subjects factor *AI accuracy* had the three levels *perfect, miss*, and *miss + false alarm*.

*4.1.4 Data analysis*

The methods used to analyse the data were mostly identical to those from Experiments 1 and 2. Exceptions will be made explicit in the respective parts of the Results section.

## 4.2 Results and discussion

*4.2.1 Ratings of AI performance*

An overview of all mean values and standard deviations is provided in Table 5.

Table 5. Means and standard deviations (in parentheses) for participants' ratings of AI performance in Experiment 3.

|  |  | Perfect | Miss | Miss + false alarm |
|---|---|---|---|---|
| AI accuracy |  | 95.8 (3.8) | 71.5 (9.8) | 65.1 (11.1) |
| Number of affected images | 2 | - | 76.5 (11.7) | 71.8 (10.2) |
|  | 3 | - | 68.4 (14.1) | 65.4 (16.7) |
|  | 4-9 | - | 74.1 (13.7) | 62.8 (13.7) |
|  | 10 | - | 66.6 (11.4) | 59.0 (13.2) |
| Number of people | 1 | 96.7 (5.2) | 61.7 (16.8) | 58.3 (17.1) |
|  | 2-4 | 94.6 (8.6) | 79.0 (10.0) | 69.9 (12.1) |
|  | 5-7 | 96.3 (6.6) | 82.2 (13.2) | 74.0 (12.9) |
|  | 8+ | 96.3 (4.6) | 66.1 (16.7) | 58.7 (18.0) |
| Position relative to the danger zone | Inside, close | - | 44.8 (29.2) | 42.0 (21.1) |
| | Inside, medium | - | 60.4 (19.4) | 60.1 (19.2) |
| | Inside, far | - | 67.7 (16.3) | 60.5 (17.4) |
| | Outside, close | - | 75.0 (10.9) | 65.0 (14.5) |
| | Outside, medium | - | 82.0 (9.5) | 74.1 (11.1) |
| | Outside, far | - | 84.9 (12.6) | 79.2 (12.5) |
| Elicitation method | Sequence-wise | 95.8 (3.8) | 71.5 (9.8) | 65.1 (11.1) |
| | Image-wise | 96.1 (3.3) | 78.7 (5.2) | 73.5 (7.0) |

*AI accuracy*. There was a highly significant effect of AI accuracy, $F(2,64) = 164.323$, $p < .001$, $\eta_p^2 = .837$. Ratings were highest in case of perfect AI (95.8), much lower for misses (71.5), and again somewhat lower for misses + false alarms (65.1), all $p$s < .001 (see Figure 3C). Numerically, the drop between misses and misses + false alarms (6.4) was much lower than the drop for false alarms in Experiment 2 (21.2). Thus, the combined cost of both mistakes does not reach up to the sum of their individual costs.

*Number of affected images*. The ANOVA revealed main effects of AI accuracy, $F(1,32) = 20.524$, $p < .001$, $\eta_p^2 = .391$, and number of affected images, $F(3,96) = 20.755$, $p < .001$, $\eta_p^2 = .393$, as well as a weak interaction of both factors, $F(3,96) = 3.078$, $p = .031$, $\eta_p^2 = .088$ (see Figure 4C). Overall, the AI was rated worse when misses occurred on more images. Yet again, only the difference between two images (74.2) and all higher numbers of images was significant, all $p$s < .002. Conversely, when more than two images were affected, it did not matter whether it was 3, 4-9, or 10 (66.9, 68.4, and 62.8, respectively), all $p$s > .1. A similar pattern was observed for misses and their combination with false alarms.

*Number of people*. There were main effects of AI accuracy, $F(2,64) = 139.787$, $p < .001$, $\eta_p^2 = .814$, and number of people, $F(3,96) = 27.997$, $p < .001$, $\eta_p^2 = .467$, as well as an interaction of both factors, $F(6,192) = 12.933$, $p = .001$, $\eta_p^2 = .288$ (see Figure 5C). The same unusual pattern occurred as in the previous experiments, both for pure misses and their combination with false alarms: Ratings were low when only one person was present, increased with more people, all $p$s < .002, were independent of the exact number of people, all $p$s > .5, but then dropped again for large groups, $p$s < .001.

*Position relative to the danger zone*. For misses and misses + false alarms, two separate one-way repeated-measures ANOVAs were conducted with the 6-level factor *position relative to danger zone* (*inside close, inside medium, inside far, outside close, outside medium, outside far*). There was a significant effect of position relative to the danger zone, both for misses $F(2.4,77.8) = 33.078$, $p < .001$, $\eta_p^2 = .508$, and misses + false alarms, $F(3.6,115.8) = 35.455$, $p < .001$, $\eta_p^2 = .526$ (see Figure 6C). In both cases, the worst ratings were given when the AI missed people on the tracks right in front of the train (44.8 and 42.0 for pure misses and combinations, respectively). In contrast, missing people far away and outside the train's profile yielded rather positive ratings (84.9 and 79.2 for misses and combinations, respectively). The large differences again suggest that people's position was a powerful determinant of participants' ratings.

*Elicitation method*. There were main effects of AI accuracy, $F(2,64) = 188.250$, $p < .001$, $\eta_p^2 = .855$, and elicitation method, $F(1,32) = 59.518$, $p < .001$, $\eta_p^2 = .650$, as well as an interaction of both factors, $F(1.6,52.4) = 39.974$, $p = .001$, $\eta_p^2 = .555$ (see Figure 7C). Sequence-wise ratings were lower than image-wise ratings both for misses and their combination with false alarms, both $p$s < .001, and the magnitude of this difference was similar between them (7.2 and 8.4, respectively).

*4.2.2 Explanations of ratings*

The information categories contained in participants' explanations closely mirrored the patterns observed in the previous experiments (see Figure 8C and Table 3): Participants often referred to the consequences of detection (danger, relevance), detection difficulty (own perception, perceivability, occlusion) and the position of people (on tracks, distance to tracks and train). Again, statements about people's position were more frequent than statements about their behaviour or physical features. The only notable difference to the previous experiments was that more image-related comments (i.e., about image context, people, and objects) were made overall. This might be because there were more AI mistakes and thus more things to comment on.

*4.2.3 Interim summary*

Experiment 3 again replicated the findings of the previous two experiments and extended them to a situation with multiple AI mistakes. In this case, the pattern of results mirrored that for plausible misses, but ratings were generally reduced by the additional false alarm. However, this reduction was less than the effect of false alarms observed in Experiment 2. Thus, the whole reduction was less than the sum of its parts. Moreover, once again there was no evidence for an influence of mistake context (i.e., AI accuracy in other images) on the ratings for plausible misses or perfect AI performance.

# 5. General discussion

How do humans decide whether an AI for automatic train operation is doing a good job at detecting objects? Three experiments investigated what factors influence human evaluations of AI performance. To this end, participants saw image sequences of people moving in the vicinity of railway tracks, and an AI had to detect and visually highlight all people. Participants had to rate the AI's performance on a numerical scale and then verbally explain these ratings. The results revealed that ratings depended on whether and how often the AI made mistakes, how much opportunity it had to make them, how easy and how dangerous it was to make mistakes, and how ratings were elicited. The influence of these factors was highly replicable across experiments and will be discussed in the following sections.

## 5.1 AI accuracy, mistake plausibility, and detection difficulty

Not surprisingly, the AI was evaluated most favourably when it correctly detected all people. In this case, it is noteworthy that participants did not seem to award any bonus points when it was harder or more important to detect people. That is, none of the influences discussed in the following sections made any difference in perfect AI trials. For instance, evaluations were no better when fifteen people were present and detected than when it was only one. A common strategy for participants was to use the maximum point score as a default for "everything is okay" and then subtract points from it if the AI made mistakes. In consequence, participants were well able to differentiate in the negative direction but not in the positive direction. This should be considered when planning and interpreting AI evaluations. If it is desirable for evaluators to appreciate instances of outstanding AI performance, audits should provide explicit options to do so. This could be accomplished by defining a medium value of zero as a default and then asking evaluators to rate both positive and negative aspects of AI performance, or by providing separate rating scales for positive and negative aspects.

When the present study's AI did make mistakes, participants punished this quite harshly. Even when people were barely visible, missing them still led to reductions of ratings by about 30 points. A smaller, additional reduction of about 12 points was observed when misses were implausible – that is, when the missed people were easy to detect and identify for a human. This is in line with plausibility effects reported for object classification (Heuer & Breiter, 2020). A novel contribution of the present study is that these effects could be pinpointed to participants' actual evaluation outcomes on the level of individual images and sequences rather than being inferred from subjective summary reports in a post-experimental interview.

While the existence of plausibility effects has been established, their causes are far from clear. For one, the cognitive anthropomorphism hypothesis assumes that humans attribute human-like perception to AI systems (Mueller, 2020). Thus, the observed plausibility effects might be based on participants' own perceptual abilities (cf. Bos et al., 2019). For instance, if they could barely see a tiny person, they might not have expected the AI to do so, either. However, there are at least two alternative explanations. One is based on a confound of plausibility and danger. Tiny or occluded people are not just hard to see, they also are further away and thus the train might still be able to brake. In this case, plausibility effects would not be a bias (i.e., cognitive anthropomorphism) but a reasonable reaction to the actual affordances of a situation. Another confound exists between plausibility and image position within a sequence. In the present study's sequences, people were most likely to appear tiny at the beginning and larger later (i.e., after the train has approached them or they have approached the train). Thus, in case of plausible misses the AI just needed some extra time to detect a person, whereas in case of implausible misses it lost a person that it had already detected. This in itself might seem implausible to humans, which was supported by numerous statements about loss and fluctuation in participants' written explanations. That is, implausibility might at least partly have resulted from the AI's dynamic behaviour. Future studies should systematically disentangle these potential contributors to better understand the mechanisms of plausibility effects.

### 5.2 Type of AI mistake

Overall, misses were evaluated more negatively than false alarms. At first glance, this seems to be at odds with previous literature, which has often reported that false alarms are more problematic (Dixon et al., 2007; Rice & McCarley, 2011; Wickens et al., 2005). This divergence might stem from the fact that in the present study, all objects and AI mistakes were easy to detect (cf. Madhavan et al., 2006). However, there are alternative explanations, for instance that false alarms might seem less dangerous in a railway context. This assumption was supported by the finding that the ratings for misses and false alarms did not differ when the missed people or falsely identified objects were far away.

Future studies should investigate under what conditions the two types of mistake influence human evaluations of AI in similar or different ways. In this regard, it would be interesting to systematically vary the plausibility of false alarms, as it was done for misses in the present study. Some of our falsely identified objects were similar to humans (e.g., posters with people, safety vests), while others were not (e.g., parts of the railway infrastructure). One might hypothesise that false alarms are evaluated more mildly if the highlighted object is easily confusable with the actual target, in line with cognitive anthropomorphism (Mueller, 2020). On the other hand, confusability might not matter, given that in the present study, evaluations of false alarms seemed immune to almost all influencing factors.

Another interesting question for future studies is whether evaluators sometimes appreciate false alarms. In the present study, some sequences contained objects that were potentially dangerous (e.g., a pram on the tracks). Several participants explained that they had given negative ratings when the AI had not detected an object that is dangerous and thus should be taken care of. Moreover, several participants stated that objects should be identified as people when they are associated with people, attached to people, or could contain people even if none are visible. This is in line with a previous finding that train drivers rely on predictive objects to infer the presence of people (Müller & Schmidt, 2024). Thus, if predictive or dangerous objects are falsely identified as people, evaluators might prefer this to not detecting the danger. Accordingly, they might not punish the AI for particular mistakes – or even punish it for not making them. A practical consequence would be that particular mistakes might go unnoticed in AI evaluations.

### 5.3 Danger of the situation

One hypothesis of the present study had been that human evaluations do not only depend on whether people are missed but also on how dangerous this is. Indeed, a missed person's position was the factor exerting the strongest influence on participants' ratings, with differences of as much as 50 % of the point range between missing a close person in the train's profile and missing a distant person outside the profile. Along the same lines, people' position was the image-related concept most prominent in participants' explanations of their ratings. A small subset of participants even indicated that they had punished the AI for highlighting people that were obviously not in danger.

This is interesting, because the AI's sole task was to detect people. The latter was made very clear in the instruction, which explicitly stated that the AI was *not* able to evaluate danger. The fact that position (as a proxy for danger) still shaped participants' ratings to such an extent, despite better knowledge, is highly relevant for the practice of AI audits. It suggests that evaluators might not actually evaluate the AI with reference to the standards it can possibly meet but instead apply their own standards. More specifically, they might shift the task definition, not evaluating performance on the assigned subtask (e.g., person detection) but on a more global task (e.g., collision avoidance). This is in line with previous findings that AI capabilities and human expectations often are misaligned (Bach et al., 2024; Kocielnik et al., 2019).

Given that participants seemed to evaluate danger, it is interesting that they almost exclusively relied on people's position as a cue. This is not what train drivers do to assess danger (Müller & Schmidt, 2024). Instead, they use a wide variety of observable cues (e.g., people's action or body language) and inferences (e.g., people's perceptual capabilities or intentions) to predict how a situation will develop. In the present study, other observable cues were rarer than position and inferences were not mentioned at all. This divergence might be a matter of expertise. Alternatively, participants might not have expected the AI to be able to interpret such cues and draw inferences, anyway, and therefore they might have excluded them from their evaluations of AI performance.

### 5.4 Methods of eliciting human evaluations

An unexpected finding of the present study was that holistic ratings of entire sequences were consistently lower than the average ratings of their individual images. Basically, this indicates that a few bad apples can ruin an entire sequence. When evaluation outcomes change with the way they are elicited, what does this mean for practice? Should audits collect evaluations of images or sequences? Or in other words, should the integration be performed algorithmically or in the evaluator's head?

On the one hand, one might argue that decisions about braking in automatic train operation will not be based on individual images but sequences. The redundancy of analysing several consecutive images allows for a compensation of momentary glitches. Thus, if people are detected in the majority of images, missing them in a few images might not be considered overly problematic. From this perspective, the disproportionately large effect of occasional misses on sequence ratings would not be desirable. Thus, image-wise ratings might be preferred, which additionally provide a higher resolution and thereby allow evaluators to pinpoint the weaknesses of an AI to particular conditions. On the other hand, image-wise ratings increase the required time and effort by several orders of magnitude. Thus, when resources are limited, sequence-wise ratings might be preferred. The present results suggest that this cheaper way of conducting audits would not result in more lenient evaluations.

Another form of eliciting image-wise ratings was not tested in the present study: randomising image order to remove all temporal sequence information. One might argue that this is the fairer condition: When the evaluated AI component runs on individual images and processes them independently, evaluators should do the same. After all, they might only have known that a blue patch was a person because they have seen that person disappear behind a bush. More generally, removing temporal sequence information would keep evaluations free from typical human expectations like object permanence in case of occlusion or other instances of bad viewing conditions. In consequence, they would presumably evaluate the AI more mildly. However, especially in safety-critical domains, this does not seem to be desirable. Taken together, selecting an elicitation method for human evaluations of AI performance requires trade-offs and between different goal criteria such as precision, effort, and safety requirements.

## 5.5 Limitations and outlook

A number of limitations constrain the interpretability and generalisability of the present findings. For one, the stimulus set of 48 sequences was quite small and some influencing factors were confounded in these sequences. For instance, sequences with large groups tended to show people in danger (e.g., crossing the tracks). Accordingly, misses were rated quite negatively, contrary to the overall trend for ratings to increase with the number of people. Using a large and fully balanced image dataset is impossible at this point, because dynamic railway image datasets are rare (for an overview see Tagiew et al., 2025). In the future, sequences should be generated that systematically and factorially manipulate different influences. If the current availability issues persist, a feasible alternative might be to use synthetic data from simulations (D'Amico et al., 2023).

A second limitation is that the present study relied on fake AI decisions and highlights. On the positive side, this enabled strict experimental control over what is highlighted and what is not. However, using actual AI outputs could provide more realistic stimuli. For instance, saliency maps created via explainable artificial intelligence (XAI) could highlight the areas that were most important for an image classifier's decision. However, such saliency maps only are useful under very restricted conditions (Müller, 2024). Thus, concept-based explanation methods might be preferred, which can additionally reveal what the AI has seen at a particular location (Poeta et al., 2023). Especially for false alarms, it would be interesting whether and how human evaluations change when they are given the chance to understand why the AI has made a particular mistake.

A third limitation is the present study's lack of differentiation, both in terms of decisions and highlights. First, person detection was binary, so that a person was either detected or not. While simplifying the interpretation of human evaluations, this approach cannot tell us how these evaluations depend on an AI system's confidence or uncertainty. Making uncertainty transparent can improve human-AI interaction (Cau et al., 2023; Jungmann et al., 2023), although it does not always lead humans to trust an image classifier or intend to rely on it (Meyers et al., 2024). In addition to the AI's decisions, also its highlights could be more differentiated, allowing for a wider range of interesting mistakes. For instance, future studies could investigate the effects of excessive or inconsistent highlighting of background areas. Would evaluations get worse, given the same AI decisions, if highlights included large areas around a detected person? Effects of highlight specificity were found in radiological settings (Famiglini et al., 2024). However, their generalisability to automatic train operation is questionable due to a number of domain differences such as the relevance of spatial precision.

A final limitation of the present study is that it was conducted with laypeople instead of experts. It is likely that AI experts or railway domain experts would have come to very different evaluations, both compared to the present study's participants and compared to each other. First, one might argue that expert auditors would be trained in evaluating AI systems and thus immune to the biases observed in the present study. However, decades of psychological research testify that instruction and training often do not make humans immune to biases. Second, the outcomes might depend on an expert's professional background. Evaluations of one and the same technical system can differ substantially between experts from different disciplines (Schmidt & Müller, 2023). For instance, AI experts might have been more sensitive to the AI's actual capabilities, whereas railway experts might have put an even stronger focus on danger than the present study's participants. Therefore, future studies should compare AI evaluations of different relevant stakeholders in the process of developing safe and explainable AI systems for railway operations.

# Acknowledgments

This work was supported by the German Centre for Rail Traffic Research (DZSF) at the Federal Railway Authority within the project "Explainable AI for Railway Safety Evaluations (XRAISE)".